\documentclass[pdftex, twocolappendix, numberedappendix, iop, apj]{emulateapj}
\usepackage{graphicx}
\usepackage{dcolumn}
\usepackage{bm}
\usepackage{amssymb}
\usepackage{latexsym}
\usepackage{booktabs}
\usepackage[flushleft]{threeparttable}
\usepackage{amsmath}
\usepackage{multirow}
\usepackage[colorlinks=true, linkcolor=red, citecolor=blue]{hyperref}
\newcommand{\be}{\begin{equation}}
\newcommand{\ee}{\end{equation}}
\newcommand{\bq}{\begin{eqnarray}}
\newcommand{\eq}{\end{eqnarray}}
\newcommand{\lcdm}{\ensuremath{\Lambda\mathrm{CDM}}}
\newcommand{\planck}{\ensuremath{\it{Planck}}}

\begin{document}
\title{Accounting for correlations when fitting extra cosmological parameters}

\author{Y.~Huang\altaffilmark{1}, G.~E.~Addison\altaffilmark{1},  C.~L.~Bennett\altaffilmark{1}}

\email{yhuang98@jhu.edu}

\altaffiltext{1}{
Dept. of Physics \& Astronomy, The Johns Hopkins University, 3400 N. Charles St., Baltimore, MD 21218-2686
}


\begin{abstract}
Current cosmological tensions motivate investigating extensions to the standard \lcdm\ model. Additional model parameters are typically varied one or two at a time, in a series of separate tests. The purpose of this paper is to highlight that information is lost by not also examining the correlations between these additional parameters, which arise when their effects on model predictions are similar, even if the parameters are not varied simultaneously. We show how these correlations can be quantified with simulations and Markov Chain Monte Carlo (MCMC) methods. As an example, we assume that \lcdm\ is the true underlying model, and calculate the correlations expected between the phenomenological lensing amplitude parameter, $A_L$, the running of the spectral index, $n_{\rm run}$, and the primordial helium mass fraction, $Y_P$, when these parameters are varied one at a time along with the \lcdm\ parameters in fits to the \planck\ 2015 temperature power spectrum. These correlations are not small, ranging from 0.31 ($A_L-n_{\rm run}$) to $-0.93$ ($n_{\rm run}-Y_P$). We find that the values of these three parameters from the \planck\ data are consistent with \lcdm\ expectations within $0.9\sigma$ when the correlations are accounted for. This does not explain the 1.8-2.7$\sigma$ \planck\ preference for $A_L>1$, but provides an additional \lcdm{} consistency test. 
For example, if $A_L>1$ was a symptom of an underlying systematic error or some real but unknown physical effect that also produced spurious correlations with $n_{\rm run}$ or $Y_P$ our test might have revealed this. We recommend that future cosmological analyses examine correlations between additional model parameters in addition to investigating them separately, one a time.
\end{abstract}
 \maketitle

\section{Introduction}

Over the last decade, much progress has been made on putting precise constraints on cosmological parameters within the $\Lambda$ cold dark matter (\lcdm{}) model, particularly from Cosmic Microwave Background (CMB) experiments \citep[e.g.,][]{Bennett2013,planck/13:2015,sievers/etal:2013,story/etal:2013}. 
In the most recent release of \planck{} results \citep{PlanckVI}, determinations of the standard \lcdm{} parameters such as baryon density, Hubble constant and matter density have reached the percent level or below. 

Although currently there is no convincing evidence for deviations from the standard \lcdm{} model from any single experiment, tensions exist between the values of some parameters inferred from different datasets. The most severe one is the 4.4$\sigma$ disagreement between the Hubble constant measurements from the anchor-Cepheid-supernova distance ladder by the SH0ES collaboration \citep{2019arXiv190307603R} 
and from the \planck{} CMB data \citep{PlanckVI}. The measurement of $H_0$ via strong lensing time delays \citep{2017MNRAS.465.4914B,birrer/etal:2019} is consistent with the distance ladder and in 2.5$\sigma$ tension with \planck{}. \cite{addison2017} showed that the tension between early and late time universe measurements persists even without the inclusion of \planck{} data, using baryon acoustic oscillation (BAO) scale measurements.

Extensions or alternatives to the \lcdm{} model have been explored in an attempt to resolve the Hubble tension. For example, the effects of varying the effective number of neutrino species \citep[e.g.,][]{Riess2016Jul} and the equation of state parameter of dark energy \citep[e.g.,][]{2017MNRAS.471.1259J} have been studied, though these extensions have not been able to effectively relieve the tension without including multiple turning points in the evolution of the dark energy equation of state \citep{zhao/etal:2017}. Recently, new ideas have been proposed as more promising solutions, for example, with the introduction of early dark energy \citep{ede2018} or self-interaction massive neutrinos \citep{2019arXiv190200534K}.

If there is new physics, consistency tests within the \lcdm{} model will eventually fail with sufficiently sensitive new data.

An example of a parameter used to test consistency is the lensing amplitude $A_L$. It was first introduced by \cite{CalabreseAL} as a phenomenological way to quantify the effect of weak gravitational lensing in the CMB power spectrum. By definition, $A_L=1$ is the physical value. However, the \planck{} temperature power spectrum (TT) data have shown a persistent preference for $A_L>1$ at 1.8-2.7$\sigma$ depending on which datasets are included \citep{planck/16:2013,planck/13:2015,PlanckVI}. 
For discussion of the $A_L$ tension, see also \cite{addison2016,motloch/hu:2018,motloch/hu:2019}.
The cause of the deviation of $A_L$ from its physical value is unclear, however, varying $A_L$ is an example of the sort of test that might ultimately shed light on the origin of the distance ladder tension.

\begin{figure}
\includegraphics[width=8cm]{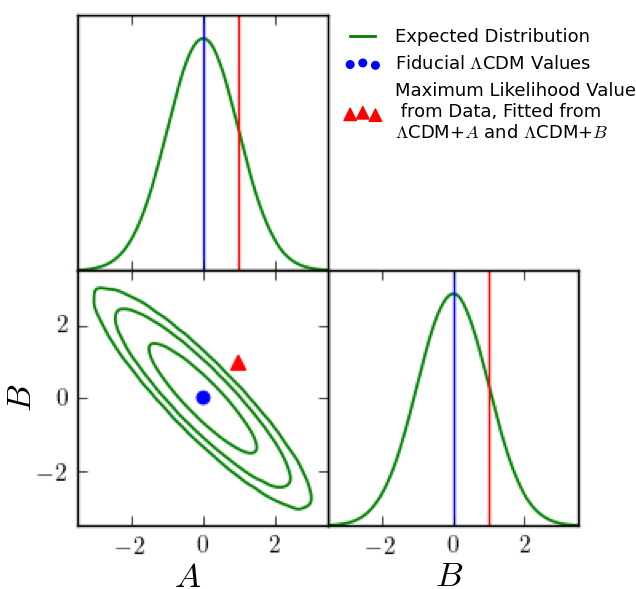}

\caption{
Accounting for correlations between additional model parameters can be important. In this example, $A$ and $B$ are additional parameters to the standard \lcdm{} model. 
The maximum likelihood (ML) values of $A$ and $B$ from \lcdm{}+$A$ and \lcdm{}+$B$ fits to a particular dataset are shown in red. The green contours show the expected distribution of ML values, calculated assuming \lcdm{}, with $A=0$ and $B=0$, is the true model. The ML values of $A$ and $B$ fitted from data appear consistent with their fiducial values in their 1-D marginalized distributions, but they are actually outside the 99.7\% contour in their 2-D distribution, suggesting a failure in the assumed model or some unknown systematics in the data.\label{fig:tri_eg}}



\end{figure}

Typically results from fitting additional model parameters are presented one or two at a time, a series of separate tests \citep[for recent examples, see][]{2017PhRvL.119j1301H, 2017MNRAS.471.1259J,PlanckVI}.  
Different parameters may have similar effects on the theory prediction 
so the constraints on extension parameters, even when fitted separately, from an experiment or particular combination of experiments may be correlated. As an example, Figure \ref{fig:tri_eg} shows the expected 2-D confidence contours for two parameters, $A$ and $B$, obtained from \lcdm$+A$ and \lcdm$+B$ fits. The green contours show the expected distribution (estimated from e.g. simulations) if \lcdm{}, with $A=0$ and $B=0$, is the true model. 
Only looking at their 1-D marginalized distributions would lead to an incorrect conclusion of consistency, as the values of $A$ and $B$ are each only 1 standard deviation away from their fiducial \lcdm{} values, while in the 2-D space the shift from the fiducial point is orthogonal to the degeneracy direction and the values are actually outside the 99.7\% contour. 
This result in the 2-D space would not mean that there is strong evidence for the \lcdm{}+$A$ or the \lcdm{}+$B$ model, since neither $A$ or $B$ disagrees with their fiducial values significantly in 1-D. Rather, it suggests that there is something else wrong with the assumed model, or there may be experimental systematics we have not accounted for.


Therefore, to more carefully assess whether the standard \lcdm{} model can consistently describe data, 
the covariance of the set of extension parameters should be quantified.

In this paper, our goal is to answer the following questions:
\begin{enumerate}
\item How do we calculate the expected correlation between different extension parameters to the standard \lcdm\ model when constrained by the same data? 
\item How do we incorporate these correlations into a more stringent test of the \lcdm\ model?
\end{enumerate}

As an example, we use the \planck{} 2015 temperature power spectrum likelihood code \citep{planck/11:2015}. The outline of this paper is as follows. In Section 2 we present the theoretical basis of our work and two different methods to achieve our goals. We show results in Section 3, followed by a discussion in Section 4 of general recipes for similar analysis in the future and conclusions in Section 5.

\section{Methodology}

\subsection{Estimating correlation between extension parameters using simulations}

In general it is not straightforward to infer the correlation between parameters $A$ and $B$ directly from MCMC fits to the \lcdm$+A$ and \lcdm$+B$ models that have already been performed by experimental collaborations like \planck. To make progress, we assume that the data likelihood is Gaussian, with a covariance matrix $\bm{\mathcal{C}}$ that does not depend on the cosmological parameters, and that the posterior distribution for the cosmological parameters is also Gaussian. 

The assumption of Gaussianity of the likelihood is made widely across different cosmological measurements, including CMB power spectra \citep[e.g.,][]{planck/11:2015,louis/etal:2017,henning/etal:2018}, weak lensing shear \citep[e.g.,][]{krause/etal:prep,hikage/etal:prep,wright/etal:prep}, galaxy clustering  \citep[e.g.,][]{percival/etal:2014,bao2017}, supernovae distance moduli \citep[e.g.,][]{scolnic/etal:2018,des/sne:prep}, and others. The likelihood can almost always be made more Gaussian through compression of the data, often with negligible loss of cosmological information (e.g., combining CMB power spectra over a range of multipoles into bins). Neglecting the cosmological parameter dependence in the covariance (e.g., assuming a fixed fiducial model and set of parameters for computing the cosmic variance contribution to the errors) has also been demonstrated to be a suitable approximation for \planck, and other current experiments \citep[e.g.,][]{hamimeche/lewis:2008,krause/etal:prep}. The fiducial model as well as the covariance are often estimated from some iterative process of fitting the actual data.

For the \planck\ data, cosmological parameter posterior distributions can be well approximated as Gaussian for \lcdm, as well as for many one- and two-parameter extensions, although there are also cases (e.g., involving curvature or varying the dark energy equation of state), where \planck\ alone does not provide Gaussian posteriors. Other experiments only constrain a portion of the \lcdm\ parameter space sufficiently well to produce Gaussian constraints in one or two parameters. Many parameters are also physically required to be positive, truncating the available parameter space and causing departure from Gaussianity when the data do not constrain the parameter to significantly differ from zero. This is the case for current cosmological constraints on the neutrino mass, for example \citep[see Figure 30 of ][]{planck/13:2015}. We return to the handling of non-Gaussian cases in Section~4.3.


In certain cases, the Bayesian posterior parameter distribution and the distribution of the maximum likelihood (ML) parameter values from many realizations of the data are approximately equal. Specifically, 
$\mathbb{P}(\bm{\theta}|\bm{d})$, the Bayesian posterior distribution of parameters $\bm{\theta}$ sampled by MCMC given the experimental data, can well approximate $\mathbb{P}(\bm{\theta}^{\rm ML}(\bm{d}^{\rm{sim}})|\bm{\theta}^{\rm{fid}})$. The latter is the distribution of frequentist maximum-likelihood (ML) parameter estimation based on realizations of $\bm{d}^{\rm{sim}}$ from a fiducial model $\bm{\theta}^{\rm{fid}}$. The choice of $\bm{\theta}^{\rm{fid}}$ is usually physically motivated and is based on fits from actual data. For a detailed discussion, see e.g., Chapter~4 and Appendix~B of \cite{gelman2013bayesian}.

We show in Appendix A that this correspondence is mathematically exact, if we: (1) assume the Gaussianity of both the data likelihood and the posterior distribution of parameters; (2) impose parameter priors that are far less constraining than the likelihood; 
(3) in the MCMC computation, replace the experimental data $\bm{d}$ with $\bm{\mu}(\bm{\theta}^{\rm{fid}})$, the theory prediction of the fiducial model. In other words, $\mathbb{P}(\bm{\theta}|\bm{d}=\bm{\mu}(\bm{\theta}^{\rm{fid}}))$ from MCMC and $\mathbb{P}(\bm{\theta}^{\rm ML}(\bm{d}^{\rm{sim}})|\bm{\theta}^{\rm{fid}})$ from simulations are the same mathematically. In this work, the fiducial model is from fitting the \planck{} TT data. We will show in Section 3.2 that the exact choice of $\bm{\theta}^{\rm{fid}}$ is not very important. 

This correspondence allows us to estimate the correlation between extension parameters $A$ and $B$ using simulated data (i.e., frequency sampling of the likelihood)
in the following steps:
\begin{enumerate}
\item Generate many simulated data sets (in our case simulated \planck-like CMB power spectra) drawn from the likelihood in the form of a Gaussian distribution $\mathcal{N}(\bm{\mu}(\bm{\theta}^{\rm{fid}}),\bm{\mathcal{C}})$ for some choice of fiducial \lcdm\ model parameters, $\bm{\theta}^{\rm fid}$, and using the covariance matrix $\bm{\mathcal{C}}$ provided by the experiment collaboration.
\item Calculate the maximum-likelihood parameters $\bm{\theta}^{\rm{ML}}$ for the \lcdm$+A$ and \lcdm$+B$ models for each simulated data set, and
\item Estimate the covariance between $A$ in the \lcdm$+A$ fit and $B$ in the \lcdm$+B$ fit using the sample covariance from the simulations.
\end{enumerate}

\subsection{Estimating correlation between extension parameters using MCMC chains}

Alternatively, in the special case of Gaussian parameter posteriors, we can run MCMC chains to estimate the correlation between extension parameters directly. In Appendix~B we show that all the information on the correlation between extension parameters $A$ and $B$ can be estimated from three sets of chains: \lcdm$+A$, \lcdm$+B$, and \lcdm$+A+B$. Specifically, we found that the correlation between $A$ and $B$ when they are fitted separately is equal to minus one times the correlation between $A$ and $B$ when they are fitted together. We refer to this property as Correlation Equivalence.

A complication of estimating correlation between extension parameters by performing MCMC on the real data in this way is that we clearly do not have control over the `underlying' cosmological model in the way we do when generating simulations. It is also possible that the real data have imperfections or systematic biases that are not correctly accounted for in the likelihood. We therefore instead performed MCMC computations replacing the real data vector with a theory prediction computed using the same $\bm{\theta}^{\rm fid}$ as for the simulations.


\subsection{An example using \planck\ CMB spectra}

Here we provide a worked example using \planck\ data for both the simulation and MCMC approaches discussed above. At the time of writing, the \planck{} 2018 likelihood code is not yet available. Therefore we perform a simple three-parameter test using the \planck{} $\ell \geq 30$ 2015 TT data from the \texttt{plik\_lite} likelihood\footnote{Can be downloaded from \url{http://pla.esac.esa.int/pla/\#cosmology}} \citep[as described in][]{planck/11:2015}. 
This simplified likelihood includes only CMB information, marginalizing over foreground template amplitudes and other nuisance parameters. 

\subsection{Fiducial model and extension parameters}
We assume the fiducial model to be the best-fit \lcdm{} model of $\ell \geq 30$ \planck{} TT \texttt{plik\_lite} data with the optical depth fixed at $\tau =0.07$, the \planck{} calibration parameter \texttt{calPlanck} equal to 1 and other nuisance parameters marginalized. The $\ell < 30$ likelihood is pixel based rather than power spectrum based. For simplicity we only include the power spectrum based likelihood of the $\ell \geq 30$ data. Moreover, because the $\ell \geq 30$ TT spectrum only well constrains the parameter combination $A_s e^{-2\tau}$, $\tau$ is fixed to break the strong degeneracy with $A_s$. As for the calibration parameter \texttt{calPlanck}, a Gaussian prior with mean equal to 1 and standard deviation 0.0025 was originally imposed on it. Its variation has minimal impact on other parameters. For simplicity, \texttt{calPlanck} is fixed at 1. 

For the standard \lcdm{} parameters, we use 
\bq&&\{\Omega_bh^2, \Omega_ch^2, 100\theta_{MC},\tau, \log A_s,n_s\}^{\rm{fid}}=\nonumber\\&&\{0.02216, 0.1211, 1.0407,0.07, 3.0777,0.9601\}. \nonumber\eq
In addition, the three extension parameters of interest and their fiducial values for this example are $\{A_L,n_{\rm run}, Y_P\}^{\rm{fid}}= \{1 ,0, 0.2453\}$. As mentioned in Section 1, $A_L$ is a phenomenological parameter that artificially scales the lensing power spectrum. It is worth investigating as \planck{} has a curious preference for $A_L>1$. 
Running of the spectral index, $n_{\rm{run}}\equiv dn_s/d \ln k$, and the primordial Helium mass fraction $Y_P$ are an interesting pair of extension parameters to test because of their high correlation ($\sim 0.9$), that is, they produce similar changes in the power spectrum damping tail. Even a small deviation from their degeneracy direction in their expected 2-D distribution would be noticeable. In \lcdm{}, $Y_P$ is calculated from $\Omega_b h^2$ and the CMB temperature through big-bang nucleosynthesis (BBN) predictions \citep{planck/13:2015}. In \lcdm$+Y_P$, $Y_P$ is independent of BBN and decouples from $\Omega_b h^2$.


The fiducial parameters from fitting the \planck{} data have uncertainties due to cosmic variance and experimental noise. A slightly different set of fiducial parameters, which are still consistent with the measured power spectrum, 
may produce different results in the covariance of the extension parameters as well as the significance of their deviations from the fiducial \lcdm{} values. We found this effect to be small for the \planck{} 2015 data. 
See Section 3.2 for details.

\subsection{Simulations}
We run simulations to sample the distribution of the extension parameters.  We draw 1000 binned power spectrum samples with mean equal to the binned power spectrum of the fiducial model and covariance equal to the \texttt{plik\_lite} CMB band power covariance. We use the same binning scheme as \planck{} 2015 to convert power spectra computed by \texttt{CAMB}\footnote{\url{https://camb.info/}} to band powers. We replace the data spectrum in the \texttt{plik\_lite} likelihood with our samples, thus forming the simulated likelihoods. Next, using the ML finding algorithm in \texttt{CosmoMC} \citep{cosmomc}, setting $\tau=0.07$ and the \planck{} calibration parameter \texttt{calPlanck} = 1, we maximize the simulated likelihoods to obtain the best-fit parameters of a specific model for each realization. The models we explore are: the standard \lcdm{}, \lcdm+$A_L$, \lcdm+$n_{\rm run}$ and \lcdm+$Y_P$.

To quantify the overall shift between the values of the extension parameters estimated from the actual data and their fiducial \lcdm{} values, we calculate the $\chi^2$ and its probability to exceed (PTE) for a $\chi^2$ distribution with degrees of freedom equal to the number of extension parameters:
\be
\chi^2=(\bm{\theta}^{\rm{ML}}_{\rm{dat}}-\bm{\theta}^{\rm{fid}})^T\bm{\Sigma}^{-1}_{\rm{ex}}(\bm{\theta}^{\rm{ML}}_{\rm{dat}}-\bm{\theta}^{\rm{fid}})
\ee
where $\bm{\Sigma}_{\rm{ex}}$ is the covariance matrix for the extension parameters only, $\bm{\theta}^{\rm{fid}}$ is an array of 3 elements equal to $\{A_L,n_{\rm run}, Y_P\}^{\rm{fid}}$ and $\bm{\theta}^{\rm{ML}}_{\rm{dat}}=(1.1245, 0.00773,0.2306)$ is an array whose values are obtained from fitting models \lcdm+$A_L$, \lcdm+$n_{\rm run}$ and \lcdm+$Y_P$ respectively on the actual \planck{} \texttt{plik\_lite} TT data with $\tau$ fixed.

\subsection{MCMC}


As mentioned in Section 2.2, we can make use of Correlation Equivalence to estimate the expected covariance $\bm{\Sigma}_{\rm{ex}}$ between extension parameters fitted separately by running MCMC chains on the data likelihood, with the experimental power spectrum replaced by the fiducial one. 
Again we set $\tau=0.07$ and the \planck{} calibration parameter \texttt{calPlanck}= 1.

The diagonal elements in $\bm{\Sigma}_{\rm{ex}}$ are estimated from the variances of the specific extension parameters from running the MCMC chains with the modified \planck{} \texttt{plik\_lite} TT likelihood, on the model \lcdm+$A_L$, \lcdm+$n_{\rm run}$ and \lcdm+$Y_P$. 

For the off-diagonal elements, first we obtain the correlation between two extension parameters (again, denoting them generically as $A$ and $B$) from the MCMC runs that vary both of them in the \lcdm{}$+A+B$ model with the same dataset. Then we calculate the covariance between $A$ and $B$, given their variances as described in the last paragraph:
 \begin{eqnarray}
Cov(A,B)_{\rm{ fitted\ separately}} &= &-corr(A,B)_{\lcdm{}+A+B} \nonumber \\
\times (Var(A)_{\lcdm{}+A}&\times& Var(B)_{\lcdm{}+B})^{\frac{1}{2}}.
\end{eqnarray}

This way, we are able to estimate all of the elements in $\bm{\Sigma}_{\rm{ex}}$ without running simulations at all, and calculate the $\chi^2$ defined in Equation (1) and its PTE to quantify the shifts of the extension parameters from their fiducial values.

\begin{figure*}
\begin{center}
\includegraphics[width=12cm]{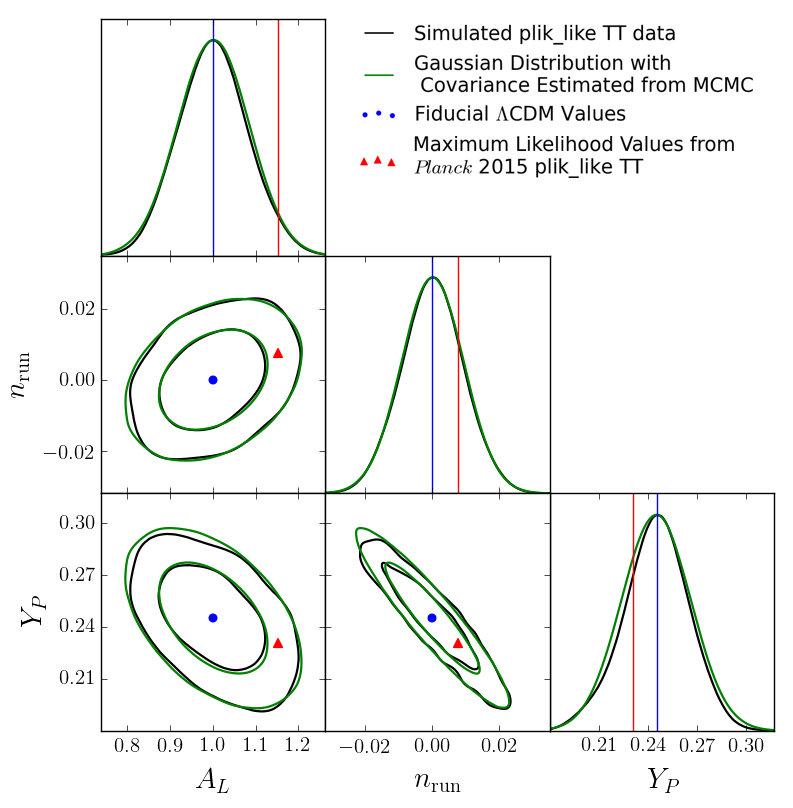}

\caption
{To illustrate the information in correlations between additional parameters fit beyond the standard \lcdm, we examine the parameters $A_L$, $n_{\rm{run}}$ and $Y_P$. Shown in red are the ML values of $A_L$, $n_{\rm run}$ and $Y_P$ in \lcdm$+A_L$, \lcdm$+n_{\rm run}$ and \lcdm$+Y_P$ fitted from the \planck{} 2015 \texttt{plik\_lite} TT data. Note that all the base \lcdm{} parameters have been marginalized over. The contours show the 68.3\% and 95.5\% confidence levels of the estimated multivariate distribution of the three parameters from simulations (black) and MCMC (green). In the black contours, there is numerical noise present due to the limited number of simulations. Notice that the ML points lie along the correlation directions, as expected from \lcdm{}. Different extension parameters may have similar effects on the predicted power spectrum. Thus their constraints from data are correlated even if they are not fitted simultaneously. Taking their correlations into account, there is no significant deviation of all three parameters from their fiducial values, which is expected if the standard \lcdm{} is the correct model.\label{fig:tri}  }

\end{center}

\end{figure*}

\begin{table}\small
\setlength\tabcolsep{6pt}
\begin{center}
\renewcommand{\arraystretch}{1.5}
\caption{ The significance of the difference between the experimental ML values and the fiducial of $A_L$, $n_{\rm{run}}$ and $Y_P$, in terms of the $\chi^2$ difference (with 3 degree of freedom) of difference, its PTE and $\Delta \sigma$, the level of consistency in terms of the number of $\sigma$.\label{chi2_table}}

\begin{tabular}{cccccccccc}

\hline\hline
\    & MCMC & simulation \\ \hline

$\chi^2$      
                   & 4.94& 4.78
                  
                   \\

PTE    
                   & 0.18  &0.19
                 
                  \\ 
$\Delta \sigma$                & 0.9 &0.9
\\
                \hline

\end{tabular}
\end{center}
\begin{tablenotes}
\small
\item 
Note. Implementing the MCMC method and running simulations give consistent results, both implying that the shifts of the experimental extension parameters from their \lcdm{} fiducial values are statistically insignificant.
\end{tablenotes}
\end{table}

\begin{table}
\setlength\tabcolsep{12pt}
\begin{center}
\renewcommand{\arraystretch}{1.5}
\caption{Parameter uncertainties and correlations for the extension parameters, estimated from the MCMC method and the simulations.\label{compare_table}}
\begin{tabular}{cccccccccc}

\hline\hline
Parameter  &MCMC& Simulations \\ \hline
\multicolumn{3}{c}{\textbf{parameter uncertainty}} \\[0.2mm]
$A_L$     &0.084  & 0.081
                  
                   \\

$n_{\rm run}$     
                   & 0.0093 &0.0093 
                 
                   \\

$Y_P$                & 0.021& 0.020  

                   \\  \hline
 \multicolumn{3}{c}{\textbf{parameter correlation}} \\[0.2mm]
$A_L$ v.s. $n_{\rm run} $       
                     & 0.31 & 0.32 
                   \\
$A_L$ v.s. $Y_P$       
                  & -0.46   & -0.47 
                   \\
$n_{\rm run}$ v.s. $Y_P$    
                    &  -0.93   & -0.93
                   \\
\hline

\end{tabular}
\end{center}
\begin{tablenotes}
\small
\item 
Note. Results from the two methods are consistent to a few percents. The largest discrepancy is in the uncertainty of $Y_P$, with 5\% difference.
\end{tablenotes}

\end{table}

\section{Results}

\subsection{Quantifying significance of deviations of extension parameters from their fiducial values}
In Table \ref{chi2_table}, we show the $\chi^2$ of the difference, as defined in Equation (1), their corresponding PTE values and the level of consistency with the assumed model in terms of the number of $\sigma$. Our results imply no significant discrepancy (0.9$\sigma$) between the values of the three additional parameters estimated from the actual data in one-parameter extensions and their fiducial \lcdm{} values. This is consistent with expectations if the standard \lcdm{} is the correct model.

Taking a closer look at elements in the estimated covariance, we show in Table \ref{compare_table} the estimated uncertainties of one-parameter extensions to the base \lcdm{} model, along with the estimated parameter correlations from the two methods. These results are consistent up to at most 5\%, in the $Y_P$ uncertainty. The differences between using the two methods are likely due to numerical noise and have only a minimal impact on our main conclusion.

We also visualize the shifts of the ML extension parameters from their fiducial values compared to their expected distributions in Figure \ref{fig:tri}. Notice in the 2-D contour plots, how the ML points lie along the correlation/degeneracy direction of the estimated distributions, as expected if the base \lcdm{} is the true model. We do not attempt to explain the 1.8$\sigma$ (from the MCMC method) or 1.9$\sigma$ (from simulations) difference between the value of $A_L$ inferred by the \texttt{plik\_lite} TT data and its physical value $A_L=1$. Rather we note that there is no extra sign of discrepancy when we take into account its non-negligible correlation with $n_{\rm{run}}$ or $Y_P$: 0.31 and -0.46 respectively. Had we found inconsistency between the ML parameters from the data and the expected \lcdm{} distribution, it could indicate the presence of systematic error or unknown physical effects producing spurious parameter correlations.



In addition, numerical differences between the  simulated contours and those of MCMC are small in this figure, again providing confidence to both methods.

\subsection{Testing stability of results against uncertainties in fiducial model}
As mentioned in Section 2.4, the fiducial parameters are from fitting the actual \texttt{plik\_lite} data and therefore have uncertainties. So the standard \lcdm{} model we define is not just one point in the parameter space, but an ensemble of points whose shape is described by the parameter posterior distribution inferred from the data.

To assess the impact of different fiducial values on our results, we randomly draw 1000 sets of parameters from the MCMC chains fitted to the \texttt{plik\_lite} data, with the base \lcdm{} model, $\tau=0.07$ and \texttt{calPlanck} = 1. For each set of parameters, we approximate the parameter covariance matrix by computing the $C_\ell$ derivatives and Fisher matrices (see Equation (B3) in Appendix B). Then we use Correlation Equivalence to calculate the extension parameter covariance $\bm{\Sigma}_{\rm{ex}}$. We find that 
varying parameters causes less than 3\% scatter in the matrix elements.

We also plug $\bm{\Sigma}_{\rm{ex}}$ into Equation (1) and find that the resulting $\chi^2$ ranges from 4.2 to 5.1 and the PTE from 0.17 to 0.24, corresponding to consistency within 0.7-1$\sigma$. These results show that the uncertainties in the fiducial model have very minimal impact on results from our consistency test. The stability of the results reflects the constraining power of the \planck{} data on the model parameters in \lcdm.

\section{Discussion}
In Sections 2 and 3 we have shown an example of quantifying the level of consistency between extension parameters and their fiducial values given a specific dataset. In this section, we outline and discuss steps to perform this approach more comprehensively for the \planck{} 2018 and other cosmology data sets.


\subsection{MCMC method}
In the ideal case where all extension parameters of interest are Gaussian, we can estimate their expected distribution from MCMC chains. We denote the individual extension parameter as ${\theta}_{\rm{ex},i}$ with $i$ runs over 1 to $n_{\rm{ex}}$, the total number of extension parameters. The suggested recipe is as follows:
\begin{enumerate}
\item Define a fiducial model (e.g. a fit from existing data) and calculate its theory prediction.
\item In the data likelihood of interest, substitute the experimental observables, e.g. power spectra in the CMB case, by the fiducial prediction.
\item Explore parameter space around the fiducial point by running MCMC chains on the modified data likelihood(s), fitting models \lcdm$+{\theta}_{\rm{ex},i}$ and \lcdm$+{\theta}_{\rm{ex},i}+{\theta}_{\rm{ex},j\neq i}$. 
\item Calculate the variance of ${\theta}_{\rm{ex},i}$ from the one-parameter extension, and the correlation between ${\theta}_{\rm{ex},i}$ and ${\theta}_{\rm{ex},j\neq i}$ from the two-parameter extension. 
\item Using results from previous step, construct a covariance matrix $\bm{\Sigma}_{\rm{ex}}$ for $\bm{\theta}_{\rm{ex}}$, with the signs of all the correlations flipped. 
\item 
Calculate $\chi^2$ defined in Equation (1) and its PTE to quantify deviation of experimental values to fiducial ones. Additionally, one can plot confidence ellipses such as in Figure \ref{fig:tri} to visualize the deviations. 
\item Check validity of the input fiducial model, for example by using a Fisher Forecast \citep[e.g.][]{2009arXiv0906.0664H} to approximate parameter covariance and estimating the shifts in the $\chi^2$ when varying fiducial parameter values.
\end{enumerate}




As an example, on a computer cluster with 12-core 2.5 GHz processors\footnote{Our computation was conducted on the computer cluster of the Maryland Advanced Research Computing Center. See \url{https://www.marcc.jhu.edu/cyberinfrastructure/hardware/} for descriptions of its system architecture.}, one MCMC run with 8 parallel chains usually take $\lesssim$ 24 hours to converge for the 2015 \texttt{plik\_lite} likelihood. The CPU time for running one MCMC job is at the order of 100 hours and the total computing time is $\sim 5\times 10^4$ hours, for running all one-parameter and two-parameter extension fits for e.g. $n_{\rm{ex}}=11$, which is the number of extra parameters fitted in the publicly available 2015 \planck{} chains.


\subsection{Simulations}
We can use simulations, as an alternative method to estimate the expected distribution of additional parameters around the fiducial point. As described in Section 2.1 and 2.5, one can perform the following procedure:
\begin{enumerate}
\item Generate simulated observables of the fiducial model using data covariance.
\item For each simulation, estimate the best-fit \lcdm+$\bm{\theta}^{ex}_i$ model for each extension parameter. 
\item Calculate the covariance of extension parameter $\bm{\Sigma}_{\rm{ex}}$, the $\chi^2$ of difference defined in Equation (1) and its PTE to quantify the significance of difference. Confidence ellipses can also be plotted.
\end{enumerate}

For a rough estimate of computing time for the simulation method, using \texttt{CosmoMC} and the same computer cluster with 12 cores per CPU, we find that the running time of the best-fit-finding algorithm is approximately 1 hour.  For $n_{\mathrm{ex}}=11$, $n_\mathrm{sim}=2000$ and running 4 parallel best-fit-finding jobs for each simulation (to reduce numerical noise and avoid obtaining results from local minima), the total computing time is $9\times10^4$ hours.


\subsection{In the case of non-Gaussianity}
To discuss the case of non-Gaussianity, first we need to clarify what exactly non-Gaussianity arises from. Recall that throughout this paper, we assume a Gaussian data likelihood and a prior that is approximately flat. If the data constrains the parameters well enough, changes in the model predictions can be treated as only linearly dependent on the parameters. This means the Taylor expansion of the log likelihood around the maximum point is significant up to quadratic terms of the parameters and therefore the parameters are Gaussian (see Appendix A).


In short, non-Gaussianity of parameters is a result of the data not being constraining enough for fitting parameters. When this is the case, there may not be a mathematical correspondence between the ML parameter distribution from the simulations and the posterior distribution of the MCMC chains, since our argument in Appendix A depends on the assumption of Gaussianity. Besides, since our proof of Correlation Equivalence in Appendix B also rests upon approximations of the log likelihood to only the second order derivatives, the property now becomes questionable. This means that we cannot simply estimate the correlation between parameters fit separately from the MCMC chains where they are fitted together. What is worse is that the means and the covariance matrix no longer carry all information of the parameter distributions and one might need to evaluate high order tensors \citep{Sellentin2014}. Additionally, the $\chi^2$ test is inappropriate for non-Gaussian parameters and we need new ways \citep[such as one proposed in Appendix C of ][]{2019PhRvD..99b3506M} to quantify the significance of the overall difference between the experimental parameters and their fiducial values.

An example of non-Gaussian parameters are those with priors that are more informative than the data, e.g. the neutrino mass $\sum m_\nu$ and the tensor to scalar ratio $r$, physically defined as non-negative. Current CMB data are not sufficiently constraining to pull these parameters away from lower bounds \citep{bicep/keck:2018,PlanckVI}.
Another non-Gaussian parameter given just the TT data, is the curvature density $\Omega_k$. Curvarture can only be weakly constrained as allowing it to be free worsens the existing degeneracy between the physical matter density $\Omega_m$ and the Hubble constant $H_0$ \citep[]{Zaldarriaga1997paramdegen,Percival2002paramdegen,2018josh}. In the MCMC method, a two-parameter extension model with both $A_L$ and $\Omega_k$ results in very wide and non-Gaussian distributions, since they are highly correlated --- a positive curvature has similar effect as an increased lensing signal \citep{planck/13:2015}.

To reduce non-Gaussianity, one can always include more datasets if available to have greater constraining power on the parameter, e.g. include the BAO data \citep{bao2017} in parameter fitting along with \planck.  However, when extensions are used as a means to test the internal consistency of one dataset, adding extra data is not an option. Fortunately, there are existing methods of transforming non-Gaussian parameters into Gaussian ones. One such method is the Box-Cox transformation \citep{Box64ananalysis}. \cite{2011MNRAS.416.1010J} and \cite{2016MNRAS.459.1916S} applied this bijective transformation to non-Gaussian cosmological parameters. Theoretically, Correlation Equivalence still holds for the transformed Gaussian parameters. 
Thus we may still apply the procedure outlined in Section 4.1 for transformed parameters, obtain the expected distribution for transformed ones, and then transform them back to the model parameters. One thing to note is that the Box-Cox transformation does not guarantee Gaussianity. So one should check for the Gaussianity of transformed parameters, e.g. calculate the skewness and kurtosis of the resulting distribution, and if needed, apply a second transformation. For consistency, it is also a good idea to compare the resulting distribution of model parameters to simulations, or compare the covariance of the transformed parameters with predictions from Fisher Forecast, keeping in mind that Fisher Forecast assumes Gaussianity.

Further understanding of non-Gaussian scenarios is left for future efforts.

\section{conclusions}


In this paper, we have presented a method to help identify potential new physics and/or systematic errors by calculating the correlations between additional parameters and performing a \lcdm{} consistency test accounting for them.

 



Usually extension parameters are added to the model separately, one a time. However, different parameters may affect the theory prediction in a similar way, which means their values from a given data set can be correlated, even when they are fit separately. Examining the consistency of extension parameters with \lcdm{} expectations, accounting for these correlations, provides an additional test of the model beyond looking at the results from individual one-parameter extensions.

Under the assumption of Gaussianity of both the likelihood and the posterior distribution of the parameters, one can fit a series of one-parameter extension models to simulated data and obtain the multivariate distribution of the extension parameters. With the base parameters marginalized over, the $\chi^2$ of difference and its PTE can be calculated to quantify the significance of deviations from \lcdm{}.

A more computationally economic alternative is to run MCMC, fitting the same series of one-parameter extension models and additionally two-parameter extensions across all the possible pairs in the set of parameters of interest. Using Correlation Equivalence as proven in Appendix B, the covariance matrix of the extension parameters fitted separately can be estimated from the results of these MCMC runs and so the expected multivariate Gaussian distribution can be obtained.

In an attempt to narrow down causes of the $A_L$ anomaly in the \planck{} data and possibly shed light on tensions between cosmological measurements, we looked at the example parameter combination $\{A_L,n_{\rm run}, Y_P\}$, fitted with the \planck{} 2015 \texttt{plik\_lite} $\ell \geq 30$ TT data, as we do not yet have access to the \planck{} 2018 likelihood. Results from MCMC and simulations show that the deviations of the three additional parameters from their fiducial \lcdm{} values are consistent with statistical fluctuations within 0.9$\sigma$ when correlations are accounted for. 

Although the cause of the reported 1.8-2.7$\sigma$ preference (depending on the the specific combination of datasets) for $A_L>1$ by the \planck{} CMB data  is yet to be understood, we find no further evidence for discrepancy when considering the correlations between $A_L$, $n_{\rm{run}}$ and $Y_P$. This is not a trivial test, as the correlations are significant: approximately 0.31 for $A_L$-$n_{\rm{run}}$, -0.46 for $A_L$-$Y_P$, and -0.93 for $n_{\rm{run}}$-$Y_P$. If the unphysical $A_L>1$ is a symptom of an underlying systematic error or some real but unknown physical effect that also produced spurious correlations with $n_{\rm run}$ or $Y_P$ our test could have revealed this.

We also tested the stability of results against the uncertainties in the parameters of the experimentally fitted fiducial model, and find that the change of the fiducial model has no impact on our conclusions, only shifting the PTE values from 0.17 to 0.24.

Furthermore, we discussed how our procedures depend on the assumption of Gaussianity of parameters. If the assumption is not valid, MCMC runs cannot simply be used to estimate correlations between parameter fitted separately, nor may there be a mathematical correspondence between parameter distributions from the simulations and the MCMC runs. 
Therefore, efforts might attempt to include Gaussianization of non-Gaussian parameters, such as using the Box-Cox transformation \citep{Box64ananalysis}. 

The procedures developed in this paper can and should be applied to more extensive lists of extension parameters, with existing and future cosmological data, in order to provide a more stringent test and complete view of \lcdm{} consistency.

\section*{Acknowledgments}
This research was supported in part by NASA grants NNX16AF28G and NNX17AF34G. This work was partly based on observations obtained with \planck\ (\url{http://www.esa.int/Planck}), an ESA science mission with instruments and contributions directly funded by ESA Member States, NASA, and Canada. This research project was conducted using computational resources at the Maryland Advanced Research Computing Center (MARCC). The \texttt{GetDist} Python package was used to make Figures \ref{fig:tri_eg} and \ref{fig:tri}. We would like to thank Josh Kable for providing his code for Fisher matrix calculations, Haoyu Wang for his help with the derivation in Appendix B, and Janet Weiland for her comments on a draft of this paper.

\appendix

\section{mathematical correspondence between frequentist maximum-likelihood parameters and bayesian parameter posterior}

In this section, we will show that there is a mathematical correspondence between frequentist maximum likelihood (ML) parameters and Bayesian parameter posterior, under conditions described below. In other words, $\mathbb{P}(\bm{\theta}^{\rm ML}(\bm{d}^{\rm{sim}})|\bm{\theta}^{\rm{fid}})$, the distribution of ML parameters estimated from realizations of a fiducial model is mathematically the same as $\mathbb{P}(\bm{\theta}|\bm{d}=\bm{\mu}(\bm{\theta}^{\rm{fid}}))$, the Bayesian posterior distribution of parameters given a set of data that matches the theory prediction of the same fiducial model.

{
Given a set of fiducial parameters $\bm{\theta}^{\rm{fid}}$, we can calculate its theory prediction $\bm{\mu}(\bm{\theta}^{\rm{fid}})$. With $\bm{\mathcal{C}}$ as the data covariance estimated experimentally, we can then draw realizations of the data vector $\bm{d}$ from a multivariate Gaussian distribution $\mathcal{N}(\bm{\mu}(\bm{\theta}^{\rm{fid}}),\bm{\mathcal{C}})$
}

Up to some constants, the log probability density function of $\bm{d}$ is given by
\be
\log \mathbb{P}(\bm{d}^{\rm{sim}}|\bm{\theta}^{\rm{fid}})= -\frac{1}{2}(\bm{d}^{\rm{sim}}-\bm{\mu}(\bm{\theta}^{\rm{fid}}))^T \bm{\mathcal{C}}^{-1}(\bm{d}^{\rm{sim}}-\bm{\mu}(\bm{\theta}^{\rm{fid}})).
\ee

For a given set of simulated data $\bm{d}^{\rm{sim}}$, we can estimate the set of parameters that best describe it by maximizing the log-likelihood 
\be\log L(\bm{\theta}) \equiv \log \mathbb{P}(\bm{d}^{\rm{sim}}|\bm{\theta}). \ee

The ML parameter estimates, $\bm{\theta}^{\rm ML}$, are defined as the solutions to the simultaneous equations
\be
\frac{\partial\log\mathbb{P}(\bm{d}^{\rm{sim}}|\bm{\theta})}{\partial\theta_i}\bigg|_{\bm{\theta}=\bm{\theta}^{\rm ML}}=0
\ee
where the index $i$ runs over all parameters. Taylor expanding around $\bm{\theta}^{\rm ML}$ gives
\be
\begin{split}
\log&\mathbb{P}(\bm{d}^{\rm{sim}}|\bm{\theta})=\log\mathbb{P}(\bm{d}^{\rm{sim}}|\bm{\theta})\big|_{\bm{\theta}=\bm{\theta}^{\rm ML}}\\ &+ \frac{1}{2}\sum_{ij}(\theta_i-\theta^{\rm ML}_i)\left[\frac{\partial^2\log\mathbb{P}(\bm{d}^{\rm{sim}}|\bm{\theta})}{\partial\theta_i\partial\theta_j}\right]_{\bm{\theta}=\bm{\theta}^{\rm ML}}(\theta_j-\theta^{\rm ML}_j)
\\ &+\mathcal{O}(\bm{\theta}-\bm{\theta}^{\rm ML})^3.
\end{split}
\ee
If higher-order terms can be neglected, Equation (A4) shows that given a set of $\bm{d}$, parameters around the ML point can be approximated by a Gaussian distribution, with the covariance $\bm{\Sigma}$ given by
\bq
\bm{\Sigma}(\bm{\theta}^{\rm ML}(\bm{d}^{\rm{sim}}))_{ij}&=&\left[-\frac{\partial^2\log\mathbb{P}(\bm{d}^{\rm{sim}}|\bm{\theta})}{\partial\theta_i\partial\theta_j}\right]^{-1}_{\bm{\theta}=\bm{\theta}^{\rm ML}}
\eq
where the term inside the brackets is the Fisher matrix and $\bm{\theta}^{\rm ML}$ is written as $\bm{\theta}^{\rm ML}(\bm{d}^{\rm{sim}})$ to emphasize its dependence on $\bm{d}^{\rm{sim}}$. 

Note that the negligibility of higher order terms in (A4) means that $\bm{\Sigma}$ is approximately constant for different values of $\bm{\theta}^{\rm ML}$. For simplicity we can set it to be $\bm{\Sigma}(\bm{\theta}^{\rm{fid}})$.

Furthermore, assuming that the zeroth order term $\log\mathbb{P}(\bm{d}^{\rm{sim}}|\bm{\theta})\big|_{\bm{\theta}=\bm{\theta}^{\rm ML}}$ is also approximately a constant for all $\bm{\theta}^{\rm ML}$, we can interchange the positions of $\bm{\theta}$ and $\bm{\theta}^{\rm ML}$ in (A4) and set $\bm{\theta}=\bm{\theta}^{\rm{fid}}$, which gives $\mathbb{P}(\bm{\theta}^{\rm ML}(\bm{d}^{\rm{sim}})|\bm{\theta}^{\rm{fid}})$, the distribution of $\bm{\theta}^{\rm ML}$ from all simulated data $\bm{d}$. Up to some constants, the log of $\mathbb{P}(\bm{\theta}^{\rm ML}(\bm{d}^{\rm{sim}})|\bm{\theta}^{\rm{fid}})$ is:
 \bq
 &&\log\mathbb{P}(\bm{\theta}^{\rm ML}(\bm{d}^{\rm{sim}})|\bm{\theta}^{\rm{fid}})=\nonumber\\&&-\frac{1}{2}(\bm{\theta}^{\rm ML}(\bm{d}^{\rm{sim}})-{\bm{\theta}}^{\rm{fid}})^T\bm{\Sigma}^{-1}(\bm{\theta}^{\rm ML}(\bm{d}^{\rm{sim}})-{\bm{\theta}}^{\rm{fid}})
 \eq


From the Bayesian viewpoint, given a single set of data $\bm{d}$, the posterior parameter distribution is given by Bayes' theorem,
\be
 \mathbb{P}(\bm{\theta}|\bm{d})=\frac{\mathbb{P}(\bm{d}|\bm{\theta})\mathbb{P}(\bm{\theta})}{ \mathbb{P}(\bm{d})}\propto\mathbb{P}(\bm{d}|\bm{\theta})\mathbb{P}(\bm{\theta}),
\ee
where $\mathbb{P}(\bm{\theta})$ is the prior probability and $\mathbb{P}(\bm{d})$ is the evidence. If the posterior is Gaussian we can write, again up to some constants,
\be
\log \mathbb{P}(\bm{\theta}|\bm{d})=-\frac{1}{2}(\bm{\theta}-\bar{\bm{\theta}})^T\bm{\Sigma}^{-1}(\bm{\theta}-\bar{\bm{\theta}}),
\ee
where $\bm{\Sigma}$ is the parameter covariance matrix given by (A5). For flat priors $\bar{\bm{\theta}}=\bm{\theta}^{\rm ML}(\bm{d})$. If $\bm{\theta}^{\rm fid}$ mentioned above is close to $\bm{\theta}^{\rm ML}(\bm{d})$, then (A6) and (A8) is only different by a small offset. To make them exactly equal, we can choose to replace $\bm{d}$ with a theory prediction computed using $\bm{\theta}^{\rm{fid}}$ so that $\bm{\theta}^{\rm ML}$ becomes equal to $\bm{\theta}^{\rm{fid}}$ and the Bayesian posterior now describes the distribution of parameters around the fiducial value:
\bq
\hspace*{-0.6cm}   
\mathbb{P}(\bm{\theta}|\bm{d}=\bm{\mu}(\bm{\theta}^{\rm{fid}}))&=&\log \mathbb{P}(\bm{\theta}|\bm{\theta}^{\rm{fid}})\nonumber\\
&=&-\frac{1}{2}(\bm{\theta}-\bm{\theta}^{\rm{fid}})^T\bm{\Sigma}^{-1}(\bm{\theta}-\bm{\theta}^{\rm{fid}}),
\eq
which matches exactly the distribution of the frequentist ML parameter estimates given in (A6). This is asymptotically true even for non-flat priors when the data are sufficiently constraining, and we do not discuss the exact choice of priors further here. For more information we again direct the reader to, for example, Chapter~4 of \cite{gelman2013bayesian}.


\section{Correlation Equivalence}
In this section, using the mathematical correspondence from Appendix A, we show that when using maximum likelihood estimation, the correlation between two parameters varied separately (e.g. the correlation between parameters $A$ and $B$ in model \lcdm+$A$ and  \lcdm+$B$) is the same but with an opposite sign as the correlation between the same two parameters varying together (e.g. \lcdm+$A+B$). 

\subsection{Maximum likelihood estimation and parameter covariance}



With $\mathfrak{L} \equiv \log L(\bm{\theta})$, the log likelihood of parameters given a specific sample of $\bm{d}$  can be written as



\begin{equation}
   \mathfrak{L}=-\frac{1}{2}(\bm{d}-\bm{\mu}(\bm{\theta}))^T \bm{\mathcal{C}}^{-1} (\bm{d}-\bm{\mu}(\bm{\theta}))
\end{equation}
up to some constants. $\bm{\mu}(\bm{\theta})$ is the theory prediction of the data given some set of parameters $\bm{\theta}$. For any $\bm{d}$, the goal is to find the $\bm{\theta}$ that maximize $\mathfrak{L}$. 

As in Appendix A, at maximum likelihood, the partial derivative of the log likelihood with respect to (w.r.t.) ${\theta}_i$ is zero ($i$ runs over $n$, the number of parameters). That is  
\bq\frac{\partial \mathfrak{L}}{\partial {\theta}_i}\bigg|_{\mathrm{ML}}&=&\frac{\partial \mathfrak{L}}{\partial \bm{\mu}^T}\bigg|_{\mathrm{ML}}\frac{\partial\bm{\mu}}{\partial {\theta}_i}\bigg|_{\mathrm{ML}}\nonumber\\ &=&-\frac{\partial \bm{\mu}^T}{\partial {\theta}_i}\bigg|_{\mathrm{ML}} \bm{\mathcal{C}}^{-1} (\bm{d}-\bm{\mu}(\bm{\bm{\theta}}))=0. \eq

Close to the ML point in parameter space, we can Taylor expand the log likelihood to the second order in $\Delta {\theta}_i\equiv {\theta}_i-{\theta}^{\rm{ML}}_i$ as shown in Equation (A4), where ${\theta}_i$ is well described by a multivariate Gaussian distribution. The parameter means are the values that maximize the likelihood.

The parameter covariance $\bm{\Sigma}$ can be calculated as the inverse of the Fisher matrix $\mathcal{F}^{-1}$, with the Fisher matrix, estimated from MCMC chains or calculated analytically, defined as:
\bq
\mathcal{F}_{ij}&=& -\frac{\partial^2 \mathfrak{L}}{\partial {\theta}_i\partial {\theta}_j}\bigg|_{\rm{ML}}  \nonumber\\
&=& \bm{\mu}^{\rm{ML},T}_{,i}\bm{\mathcal{C}}^{-1} \bm{\mu}^{\rm{ML}}_{,j}.
 \eq
From now on we use the comma notation to denote partial derivatives w.r.t. $\bm{\theta}$, that is, $\bm{\mu}_{,i}\equiv \frac{\partial \bm{\mu}}{\partial \theta_i}$.

In the following proof, we continue to work under the assumption of Gaussianity for $\bm{\theta}$ and expand $\bm{\mu}$ only to the first order of $\bm{\theta}$, that is, assuming the Gaussian linear model \citep{2018arXiv180604649R}, around a chosen fiducial model $\bm{\theta}^{\mathrm{fid}}$:
\begin{equation} \bm{\mu}(\bm{\theta}) \approx \bm{\mu}^{\mathrm{fid}}+\sum^n_{i=1} \bm{\mu}^{\mathrm{fid}}_{,i}({\theta}_i-{\theta}^{\mathrm{fid}}_i),
 \end{equation}
 where we define $\bm{\mu}^{\mathrm{fid}}= \bm{\mu}(\bm{\theta}^{\mathrm{fid}})$. In the Gaussian linear model, the Jacobian $\bm{\mu}_{,i}$ may be taken as constant over changes in $\bm{\theta}$ and so $\bm{\mu}^{\mathrm{fid}}_{,i}$ is approximately equal to $\bm{\mu}^{\mathrm{ML}}_{,i}$. For simplicity, from here on, we just drop all the superscripts for $\bm{\mu}_{,i}$. In practice, the fiducial model is usually based on to the ML model given by the data and updated iteratively if necessary. 

  
Next, we move on to calculate the elements in parameter covariance $\bm{\Sigma}$:
\begin{equation}
\bm{\Sigma}_{ij}=(\mathcal{F}^{-1})_{ij}.
 \end{equation}
Remember that for any invertible square matrix $\mathcal{M}$, its inverse can be calculated as 
 \begin{equation}
(\mathcal{M}^{-1})_{ij}=(-1)^{i+j}\frac{|\mathcal{M}_{\setminus j \setminus i}|}{|\mathcal{M}|}
 \end{equation}
 where $|\mathcal{M}|$ is the determinant of $\mathcal{M}$. We use the notation ``$\mathcal{M}_{\backslash j \backslash i}$",  to denote a smaller matrix, corresponding to $\mathcal{M}$ with the $j^{\rm{th}}$ row and the $i^{\rm{th}}$ column removed (the backslash symbol is borrowed from the notation for set difference in set theory). Then ${|\mathcal{M}_{\setminus j \setminus i}|}$ is a minor of $\mathcal{M}$.
 And the determinant for an $n\times n$ matrix $\mathcal{M}$ can be written as 
  \begin{equation}
  |\mathcal{M}|=\sum^{n}_{i=1} (-1)^{i+j} \mathcal{M}_{ij} |\mathcal{M}_{\backslash i\backslash j}|
  \end{equation}
 for any $1\leq j\leq n$.
 
 Therefore \begin{equation}
\bm{\Sigma}_{ij} = (-1)^{i+j}|\mathcal{F}_{\backslash j \backslash i}|/|\mathcal{F}|.
 \end{equation}
 
If we choose to fix the $i^{\rm{th}}$ parameter, we can simply remove the $i^{\rm{th}}$ row and column from the Fisher matrix, and calculate a new parameter covariance matrix from the revised Fisher matrix.

\subsection{Correlation between parameter $A$ and $B$, varying together}
When two parameters $A$ and $B$ are both variables, we can calculate their covariance from the Fisher matrix that includes them along with the base \lcdm{}  parameters. So their correlation is just
\begin{equation}
corr(A,B)= \frac{\bm{\Sigma}_{AB}}{(\bm{\Sigma}_{AA}\bm{\Sigma}_{BB})^{\frac{1}{2}}}.
 \end{equation}
 From here on, for clarity, we use $A$ and $B$ as subscripts for the rows and columns in the matrix for the two parameters of interest.
 
 Following from equation (B8), (B9) becomes
 \begin{equation}
corr(A,B) = \frac{-|\mathcal{F}_{\backslash A\backslash B}|}{(|\mathcal{F}_{\backslash A\backslash A}||\mathcal{F}_{\backslash B\backslash B}|)^{\frac{1}{2}}}.
 \end{equation}
The minus sign is from setting the index of $B$ equal to that of $A$ plus one.

\subsection{Correlation between parameter $A$ and $B$, varying separately}
 For a generic set of parameters, given the data array $\bm{d}$ (experimental or simulated) and the Fisher matrix $\mathcal{F}$, we can substitute (B5) into (B2) to obtain a relationship between the ML parameters $\bm{\theta}^{\rm{ML}}$ and the fiducial ones:
\be
\bm{\theta}^{\rm{ML}}(\bm{d})-\bm{\theta}^{\rm{fid}}=\mathcal{F}^{-1} \bm{\mu}^T_{,i}\bm{\mathcal{C}}^{-1}(\bm{d}-\bm{\mu}).
\ee 
So we can express $\bm{\theta}^{\rm{ML}}$ as:
 \begin{equation}
\bm{\theta}^{\rm{ML}}(\bm{d})=\mathcal{F}^{-1} \bm{\mu}^T_{,i}\bm{\mathcal{C}}^{-1}\bm{d}+\hat{\bm{\theta}},
 \end{equation}
 where we use the vector $\hat{\bm{\theta}}$ to represent all the terms in (B11) that are independent of data $\bm{d}$:
 \begin{equation}
\hat{\bm{\theta}}\equiv -\mathcal{F}^{-1} \bm{\mu}^T_{,i} \bm{\mathcal{C}}^{-1}\bm{\mu}^{\rm{fid}}+\bm{\theta}^{\rm{fid}}. \nonumber
 \end{equation}
As we will show below, $\hat{\bm{\theta}}$ does not contribute to the correlation between $A$ and $B$.

Again for clarity, we use $A$ and $B$ to index the rows and columns for parameter $A$ and $B$, respectively, while using the generic $i$ and $j$  to index $n$ base \lcdm{} parameters. So $A$ is the $(n+1)^{\rm{th}}$ parameter and $B$ $(n+2)^{\rm{th}}$. 

When fixing $B$ to its fiducial \lcdm{} value $B^{\mathrm{fid}}$, the expression for the best estimate for other parameters is almost the same as equation (B12), except that $\mathcal{F} \rightarrow \mathcal{F}_{\backslash B\backslash B}$ and the elements associated with $B$ in $ \bm{\mu}^T_{,i}$, $\bm{\theta}^{\rm{ML}}$ and $\hat{\bm{\theta}}$ are deleted. 
The expression for the rest of elements in $\bm{\theta}^{\rm{ML}}$ is:
 \begin{equation}
 \left ( \begin{array}{ccc}
        \bm{\theta}^{\rm{ML}}_1 \\
        \bm{\theta}^{\rm{ML}}_2 \\
        \vdots\\
        \bm{\theta}^{\rm{ML}}_n\\
        A^{\rm{ML}}
       \end{array} \right )=( \mathcal{F}_{\backslash B\backslash B})^{-1} \left ( \begin{array}{ccc}
        \bm{\mu}^T_{,1}\bm{\mathcal{C}}^{-1}\bm{d} \\
         \bm{\mu}^T_{,2}\bm{\mathcal{C}}^{-1}\bm{d} \\
        \vdots\\
         \bm{\mu}^T_{,n}\bm{\mathcal{C}}^{-1}\bm{d} \\
         \bm{\mu}^T_{,A}\bm{\mathcal{C}}^{-1}\bm{d} 
       \end{array} \right ) + \hat{\bm{\theta}}.
\end{equation}

The last row of (B13) gives
 \begin{eqnarray}
&&A^{\rm{ML}}(\bm{d})=\\&&\bigg[ \sum^n_{i=1} ( \mathcal{F}_{\backslash B\backslash B})^{-1}_{Ai}  \bm{\mu}^T_{,i}\bm{\mathcal{C}}^{-1}+  ( \mathcal{F}_{\backslash B\backslash B})^{-1}_{AA} \bm{\mu}^T_{,A}\bm{\mathcal{C}}^{-1} \bigg] \bm{d}\nonumber
+\hat{{\theta}}_A.
\end{eqnarray}

Similarly, fixing $A$ and letting $B$ vary leads to 
 \begin{eqnarray}
&&B^{\rm{ML}}(\bm{d})=\\&&\bigg[\sum^n_{i=1} ( \mathcal{F}_{\backslash A\backslash A})^{-1}_{Bi}  \bm{\mu}^T_{,i}\bm{\mathcal{C}}^{-1}+  ( \mathcal{F}_{\backslash A\backslash A})^{-1}_{BB} \bm{\mu}^T_{,B}\bm{\mathcal{C}}^{-1} \bigg]\bm{d}\nonumber
+\hat{{\theta}}_B.
\end{eqnarray}

The variance of $A$ and $B$ can be obtained from the Fisher matrix (B3) and its minors (B7), as follows:
 \begin{eqnarray}
Var(A^{\rm{ML}})&=&( \mathcal{F}_{\backslash B\backslash B})^{-1}_{AA} =\frac{\Big|\mathcal{F}_{\substack{\backslash B\backslash B \\ \backslash A\backslash A}}\Big|}{|\mathcal{F}_{\backslash B\backslash B }|} \\
Var(B^{\rm{ML}})&=&( \mathcal{F}_{\backslash A\backslash A})^{-1}_{BB} =\frac{\Big|\mathcal{F}_{\substack{\backslash A\backslash A \\ \backslash B\backslash B}}\Big|}{|\mathcal{F}_{\backslash A\backslash A }|}.
\end{eqnarray}
And the covariance between $A^{\rm{ML}}$ and $B^{\rm{ML}}$ is:
\be
Cov(A^{\rm{ML}},B^{\rm{ML}})=\langle (A^{\rm{ML}} - \langle A^{\rm{ML}}\rangle)(B^{\rm{ML}} - \langle B^{\rm{ML}}\rangle)\rangle,
\ee
with the brackets here representing the averaging over all realizations of the data $\bm{d}$ .

Recall that the data covariance $\bm{\mathcal{C}}$ is assumed to be fixed and with our assumption of a Gaussian Linear Model, all partial derivatives w.r.t. the parameters are also constant. Then in (B14) and (B15), only $\bm{d}$ is a variable. Inserting (B14) and (B15) into (B18), 
we find that terms involving $\hat{{\theta}}_A$ and $\hat{{\theta}}_B$ cancel. In addition, all constant terms can be taken out of the brackets, leaving only  $\langle (\bm{d} - \langle\bm{d}\rangle)^T(\bm{d} - \langle\bm{d}\rangle)\rangle$, which is just $\bm{\mathcal{C}}$. This simplifies to
 \begin{eqnarray}
 &&Cov(A^{\rm{ML}}, B^{\rm{ML}})=\nonumber\\&&\sum^n_{i,j=1}( \mathcal{F}_{\backslash B\backslash B})^{-1}_{Ai} \bm{\mu}^T_{,i}\bm{\mathcal{C}}^{-1} \bm{\mu}_{,j}(\mathcal{F}_{\backslash A\backslash A})^{-1}_{Bj} \nonumber \\
 &&+\sum^n_{i=1}( \mathcal{F}_{\backslash B\backslash B})^{-1}_{Ai} \bm{\mu}^T_{,i}\bm{\mathcal{C}}^{-1}\bm{\mu}_{,B}(\mathcal{F}_{\backslash A\backslash A})^{-1}_{BB} \nonumber \\
  &&+\sum^n_{i=1}( \mathcal{F}_{\backslash B\backslash B})^{-1}_{AA} \bm{\mu}^T_{,A}\bm{\mathcal{C}}^{-1}\bm{\mu}_{,i}(\mathcal{F}_{\backslash A\backslash A})^{-1}_{Bi} \nonumber \\
     &&+( \mathcal{F}_{\backslash B\backslash B})^{-1}_{AA} \bm{\mu}^T_{,A}\bm{\mathcal{C}}^{-1}\bm{\mu}_{,B}(\mathcal{F}_{\backslash A\backslash A})^{-1}_{BB}.
 \end{eqnarray}
Notice that terms like $\bm{\mu}^T_{,i}\bm{\mathcal{C}}^{-1}\bm{\mu}_{,j}$ are just elements of the Fisher matrix and we can also use equation (B7) to express terms like $( \mathcal{F}_{\backslash B\backslash B})^{-1}_{Ai}$ in terms of determinants of minors of the Fisher matrix. Thus,
\begin{eqnarray}
 &&Cov(A^{\rm{ML}},B^{\rm{ML}})= \frac{1}{|\mathcal{F}_{\backslash B\backslash B}||\mathcal{F}_{\backslash A\backslash A}|} \times\nonumber\\
& \Bigg[& \underbrace{\sum^n_{i,j=1}(-1)^{i+n+1}\Big| \mathcal{F}_{\substack{\backslash B\backslash B \\ \backslash i\backslash A}}\Big| \mathcal{F}_{ij}(-1)^{j+n+1}\Big| \mathcal{F}_{\substack{\backslash A\backslash A \\ \backslash j\backslash B}}\Big| }_{\text{[1]}}\nonumber   \\
 &+& \underbrace{\sum^n_{i=1} (-1)^{i+n+1}\Big| \mathcal{F}_{\substack{\backslash B\backslash B \\ \backslash i\backslash A}}\Big| \mathcal{F}_{iB} \Big| \mathcal{F}_{\substack{\backslash A\backslash A \\ \backslash B\backslash B}}\Big| }_{\text{[2]}} \nonumber \\
&+& \underbrace{\sum^n_{i=1} \Big| \mathcal{F}_{\substack{\backslash A\backslash A \\ \backslash B\backslash B}}\Big|  \mathcal{F}_{Ai} (-1)^{i+n+1}\Big| \mathcal{F}_{\substack{\backslash A\backslash A \\ \backslash i\backslash B}}\Big|  }_{\text{[3]}} \nonumber \\
&+& \underbrace{ \Big| \mathcal{F}_{\substack{\backslash B\backslash B \\ \backslash A\backslash A}} \Big| \mathcal{F}_{AB} \Big| \mathcal{F}_{\substack{\backslash A\backslash A \\ \backslash B\backslash B}}\Big|}_{\text{[4]}}  \Bigg].
 \end{eqnarray}

To simplify (B20), first we combine term [1] and [2]:
\begin{eqnarray}
&&[1]+[2]\nonumber\\
&=&\sum^n_{i=1} (-1)^{i+n+1}\Big| \mathcal{F}_{\substack{\backslash B\backslash B \\ \backslash i\backslash A}}\Big| \times \nonumber \\ &\Bigg[& \sum^n_{j=1} \mathcal{F}_{ij}(-1)^{j+n+1}\Big| \mathcal{F}_{\substack{\backslash A\backslash A \\ \backslash j\backslash B}}\Big| + \mathcal{F}_{iB} \Big| \mathcal{F}_{\substack{\backslash A\backslash A \\ \backslash B\backslash B}}\Big|  \Bigg] \nonumber \\
&=&\sum^n_{i=1} (-1)^{i+n+1}\Big| \mathcal{F}_{\substack{\backslash B\backslash B \\ \backslash i\backslash A}}\Big| \times \nonumber \\ &\Bigg[& \sum^n_{j=1} \mathcal{F}_{ji}(-1)^{(n+1)+j}\Big| \mathcal{F}_{\substack{\backslash A\backslash A \\ \backslash j\backslash B}}\Big| +  \mathcal{F}_{Bi} (-1)^{2\times(n+1)}\Big| \mathcal{F}_{\substack{\backslash A\backslash A \\ \backslash B\backslash B}}\Big|  \Bigg] \nonumber
 \end{eqnarray}
 where we have used the property that the Fisher matrix is symmetric.

Compared to equation (B8), note that the terms inside the last bracket above sum up to the determinant of an $(n+1)\times(n+1)$ matrix, which is the same as  $\mathcal{F}_{\backslash A\backslash A}$, except that its $(n+1)^{\rm{th}}$ column is replaced by its $i^{\rm{th}}$.  So not all of the columns for this matrix are linearly independent, resulting in its determinant being zero. Thus $[1]+[2]=0$.
 
To simplify term [3] and [4], we have
\begin{eqnarray}
&&[3]+[4]\nonumber\\ &=& \Big| \mathcal{F}_{\substack{\backslash A\backslash A \\ \backslash B\backslash B}}\Big| \Bigg[ \sum^n_{i=1}\mathcal{F}_{Ai} (-1)^{i+n+1}\Big| \mathcal{F}_{\substack{\backslash A\backslash A \\ \backslash i\backslash B}}\Big| + \mathcal{F}_{AB} \Big| \mathcal{F}_{\substack{\backslash A\backslash A \\ \backslash B\backslash B}} \Big| \Bigg]\nonumber\\
&=& \Big| \mathcal{F}_{\substack{\backslash A\backslash A \\ \backslash B\backslash B}}\Big| \Bigg[ \sum^n_{i=1}\mathcal{F}_{iA} (-1)^{i+n+1}\Big| \mathcal{F}_{\substack{\backslash A\backslash B \\ \backslash i\backslash A}}\Big| + \mathcal{F}_{BA} \Big| \mathcal{F}_{\substack{\backslash A\backslash B \\ \backslash B\backslash A}} \Big| \Bigg]\nonumber\\
&=& \Big| \mathcal{F}_{\substack{\backslash A\backslash A \\ \backslash B\backslash B}}\Big| \Big| \mathcal{F}_{\backslash A\backslash B}\Big|
 \end{eqnarray}

Insertng (B21) into (B20), we have
\begin{equation}
 Cov(A^{\rm{ML}},B^{\rm{ML}})=\frac{ \Big| \mathcal{F}_{\substack{\backslash A\backslash A \\ \backslash B\backslash B}}\Big| \Big| \mathcal{F}_{\backslash A\backslash B}\Big|}{|\mathcal{F}_{\backslash B\backslash B}||\mathcal{F}_{\backslash A\backslash A}|}.\end{equation}
 \\
Using Equation (B16) and (B17), the correlation between the best-fit values of parameter $A$ and $B$ estimated separately can be written as 
 \begin{eqnarray}
 corr(A^{\rm{ML}},B^{\rm{ML}})&=&\frac{Cov(A^{\rm{ML}},B^{\rm{ML}})}{(Var(A^{\rm{ML}})\times Var(B^{\rm{ML}}))^{\frac{1}{2}}} \nonumber \\
 &=& \frac{ \Big| \mathcal{F}_{\substack{\backslash A\backslash A \\ \backslash B\backslash B}}\Big| \Big| \mathcal{F}_{\backslash A\backslash B}\Big|}{|\mathcal{F}_{\backslash B\backslash B}||\mathcal{F}_{\backslash A\backslash A}|} \frac{(|\mathcal{F}_{\backslash B\backslash B}||\mathcal{F}_{\backslash A\backslash A}|)^{\frac{1}{2}}}{ \Big| \mathcal{F}_{\substack{\backslash A\backslash A \\ \backslash B\backslash B}}\Big|} \nonumber \\ 
  &=& \frac{|\mathcal{F}_{\backslash A\backslash B}|}{(|\mathcal{F}_{\backslash A\backslash A}||\mathcal{F}_{\backslash B\backslash B}|)^{\frac{1}{2}}},
  \end{eqnarray}
  which is the same as equation (B10) up to a minus sign, thus completing our proof.


\end{document}